# Relaxor ferroelectric behavior and structural aspects of

# $SrNaBi_2Nb_3O_{12}$ ceramics

Sunil Kumar,[1] D.A. Ochoa,[2] J.E. Garcia[2], and K.B.R. Varma,[1,a)]

[1] Materials Research Centre, Indian Institute of Science, Bangalore, India 560012.

[2] Department of Applied Physics, Universitat Politècnica de Catalunya, 08034

Barcelona, Spain.

**Abstract**

$SrNaBi_2Nb_3O_{12}$ powder was prepared via the conventional solid-state reaction method. X-ray structural studies confirmed the phase to be a three-layered member of the Aurivillius family of oxides. $SrNaBi_2Nb_3O_{12}$ (SNBN) ceramics exhibited the typical characteristics of relaxor ferroelectrics, associated with broad and dispersive dielectric maxima. The variation of temperature of dielectric maxima ($T_m$) with frequency obeyed the Vogel-Fulcher relationship. Relaxor behavior was believed to be arising from the cationic disorder at A-site. Pinched ferroelectric hysteresis loops were observed well above $T_m$.

[a)] Corresponding author; E-Mail: kbrvarma@mrc.iisc.ernet.in, Tel.: +91 80 22932914.

## I Introduction

Aurivillius family of oxides have become increasingly important as lead-free ferroelectric materials owing to their superior polarization fatigue resistance characteristics along with high Curie temperatures that make them attractive for nonvolatile memory applications.[1,2] These bismuth layer-structured ferroelectrics consist of intergrowth of $[Bi_2O_2]^{2+}$ layers and $[A_{n+1}B_nO_{3n+1}]^{2-}$ pseudo-perovskite units, where $n$ represents the number of perovskite-like layers stacked along the c-axis. Ferroelectricity in the orthorhombic phase of these compounds is generally attributed to the cationic displacement along the polar a-axis and the tilting of octahedra around the a- and c-axes.[3-5] Though the majority of Aurivillius compounds have normal ferroelectric to paraelectric phase transitions, a few compounds however exhibit relaxor-like diffuse phase transitions.[6-11] Relaxors that are characterized by the frequency dependent diffuse phase transitions, are of significant interest for various applications as these possess exceptionally high dielectric and piezoelectric responses over a wide range of temperatures.[12,13]

Most literature pertaining to relaxors constitutes the studies on lead-based complex perovskites. In comparison, comprehension of the mechanism underlying the relaxor behavior in Aurivillius compounds is limited. Positional static disorder on A-site as well as the presence of Ba cations in $Bi_2O_2$ layer associated with the composition fluctuations which result in the formation of nano/micro domains with different structural distortion levels is reckoned to be the possible mechanism for the relaxor behavior in $BaBi_2(Ta/Nb)_2O_9$ and $BaBi_4Ti_4O_{15}$.[11,13,14] Karthik et al. indeed confirmed the presence of ordered polar regions with 1:1 ordering of A-site cations and accompanied relaxor behavior in $K_{0.5}La_{0.5}Bi_2(Nb/Ta)_2O_9$.[7,8] Investigations concerning the structural, dielectric and the relaxor characteristics of $SrNaBi_2Nb_3O_{12}$ (three-layer Aurivillius phase) are reported in the present article. $SrNaBi_2Nb_3O_{12}$ ceramics exhibited Polarization-electric field (P-E) hysteresis loops at temperature well above the temperature of dielectric maximum ($T_m$). This behavior is consistent with the existence of polar nano/micro regions above $T_m$ in relaxor ferroelectrics.

## II Experimental

Strontium sodium bismuth niobate (henceforth abbreviated as SNBN) powder was synthesized by reacting stoichiometric amounts of reagent grade $Na_2CO_3$, $SrCO_3$, $Bi_2O_3$ and $Nb_2O_5$ (purity > 99%). The above mixture was ball milled using agate balls in ethanol medium for 4 h, dried and calcined at 1073 K for 5 h and subsequently at 1223 K for 10 h with intermediate grinding. The calcined powder was mixed with a small amount of PVA (polyvinyl alcohol) and the mixture was uniaxially pressed into pellets of about 1-2 mm in thickness and 12 mm in diameter. The binder was burnt out by slowly heating the pellets at 773 K for 5 h. The pressed pellets were sintered at 1323 K for 4 h in air at a heating rate of 3 K/ min and then furnace cooled to room temperature. The monophasic nature of the calcined powders and subsequently sintered pellets was confirmed by X-ray powder diffraction (Bruker-D8 Advance, Cu-$K_\alpha$ radiation) studies. The *FULLPROF* program was used for Rietveld structural refinement. The Bragg peaks were modelled with pseudo-Voigt function and the background was estimated by linear interpolation between selected back-ground points. Scale factor, zero correction, half width parameters, lattice parameters, and positional coordinates were varied during the refinement. Densities of the sintered pellets were measured by the Archimedes method. Xylene (room temperature density ~ 0.86 g/cm$^3$) was used as the liquid medium. Scanning electron microscope (Quanta-ESEM) was used for microstructural analyses. The high resolution transmission electron microscopic (HRTEM) studies were carried out using a Tecnai-T20 (200 kV) TEM equipped with a double-tilt rotating sample holder. Specimens for TEM were prepared by sonicating the dispersed SNBN powder in acetone medium and putting a droplet of this dispersion on a copper grid coated with carbon thin film. For electrical property measurements, silver electrodes were applied on either side of polished surfaces of pellet and then the sample was baked at 473 K for 2 h to dry out the moisture prior to any measurement. Dielectric response was obtained by using a precision LCR meter (Agilent E4980A) sweeping between 100 Hz and 1 MHz. Samples were placed in a closed loop cold finger cryogenic system in order to scan the temperature between 50 K and 400 K at 1 K min$^{-1}$. The ferroelectric characteristics; the remnant polarization, maximum polarization and coercive field were determined from the P–E hysteresis loop recorded using a modified Sawyer–Tower circuit (Automated PE loop tracer, Marine India Elect.).

**III Results and Discussion**

Figure 1(a) shows X-ray powder diffraction pattern, obtained at room temperature, along with the Rietveld refinement profile for SNBN. All the diffraction peaks could be indexed to the space group $B2cb$. The reliability factors $R_p$, $R_{wp}$ (weighted profile), $R_{exp}$, $\chi^2$ and the goodness of fit s (= $R_{wp}/R_{exp}$) for the pattern are 5.99%, 8.45%, 4.46%, 3.59% and 1.89, respectively. The refined lattice parameters are: $a$ =5.5122(3) Å, $b$=5.5093(3) Å, and $c$ =32.9150(12) Å. Pseudo-tetragonality at room temperature in SNBN is expected as the phase transition is in the vicinity of room temperature and microscopic polar regions are indeed present even above the phase transition temperature (discussed latter in the text). It should be mentioned here that the choice of orthorhombic $B2cb$ space group was in light of the observations that odd-layered Aurivillius compounds tend to crystallize in $B2cb$ space group in their ferroelectric state and the incidence of a ferroelectric hysteresis loop at room temperature confirm the ferroelectric nature of SNBN. Detailed structural analysis for the SNBN goes beyond the scope of the present investigation and remains to be addressed to in future work. The lattice image recorded along the [100] zone axis is shown in the Fig. 1(b). Regular stacking of the slabs at a periodicity of ~ 1.65 nm which corresponds to the half of the $c$ lattice parameter of the orthorhombic unit cell establishes the layered structure of the present compound. The scanning electron micrograph recorded for the pellet sintered at 1323 K for 4 h in air is shown in Fig. 1(c). The micrograph shows the presence of well-packed and randomly oriented platelet-like grains with an average size of ~10 μm. Energy dispersive X-ray analyses (EDX) carried out on various grains of the sample confirmed the composition to be $SrNaBi_2Nb_3O_{12}$ within the typical error associated with EDX analysis. The relative density of this pellet was around 95% of theoretical density. This disc-like morphology is indeed a characteristic feature of the Aurivillius family of compounds and is attributed to their strong anisotropic nature.[16]

Temperature dependence of the real ($\varepsilon_r'$) and imaginary ($\varepsilon_r''$) parts of the dielectric constant for the SNBN sample at selected frequencies in the range 100 Hz - 1MHz is depicted in Fig. 2. Dielectric data was collected during heating the sample with a constant heating rate of 1 K/min. Diffuse phase transition associated with strong frequency dispersion is observed in both $\varepsilon_r'$ and $\varepsilon_r''$ curves, suggesting the relaxor nature of SNBN ceramics. The magnitude of $\varepsilon_r'$ decreases with increasing frequency and the temperature of dielectric maximum $T_m$, corresponding to the maximum value of $\varepsilon_r'$, shifts towards higher temperature. A shift of 25 K in $T_m$ with increase in frequency from 100 Hz to 1 MHz is observed. Similar shift in $T_m$ and an increase in maximum value of $\varepsilon_r''$ are also observed with increasing frequencies. To quantify the diffuseness of phase transition, a modified Curie–Weiss law given by $1/\varepsilon_r' - 1/\varepsilon_m = C^{-1}(T - T_m)^\gamma$ is used;[17] where $\varepsilon_m$ is the dielectric constant at $T_m$, C is the Curie-like constant and $\gamma$ $(1 \leq \gamma \leq 2)$ is the degree of diffuseness. By fitting the experimental data in modified Curie-Weiss law, $\gamma$ is found to be $1.67 \pm 0.01$ (at 100 kHz) corroborating the diffused phase transition behavior of SNBN ceramics.

Relaxor ferroelectrics are known to follow an empirical Vogel–Fulcher (VF) relationship,[18] $f=f_o\exp[-E_a/k(T_m-T_{VF})]$, where $E_a$ and $T_{VF}$ are the activation energy and the static freezing temperature, $k$ is the Boltzmann constant and $f_o$ is the pre-exponential factor. Figure 3 shows the variation of $T_m$ with $\ln(f)$ for the SNBN ceramics. From the non-linear VF fitting of $\varepsilon_r'$ [Fig. 3], values of the fitting parameters $f_o$, $E_a$ and $T_{VF}$ were found to be $1.1 \times 10^8$ Hz, $16.2 \pm 0.7$ meV and $281.1 \pm 0.6$ K, respectively. The values of activation energy and static freezing temperature for SNBN are comparable to that of conventional lead-based relaxor $Pb(Mg_{1/3}Nb_{2/3})O_3$ $(T_{VF} \sim 291$ K and $E_a \sim 40$ meV).[18] Comparison with the other relaxors belonging to Aurivillius family of oxides shows that $T_{VF}$ for the present compound is significantly lower than that for $BaBi_4Ti_4O_{15}$[19] and $K_{0.5}La_{0.5}Bi_2(Nb/Ta)_2O_9$.[7,8] and higher than the static freezing temperature reported for $BaBi_2Nb_2O_9$.[6] The activation energy of frequency dispersion for $BaBi_2Nb_2O_9$ is about 0.6 eV and $T_m-T_{VF}$ is about 300 K (at 1 kHz). In comparison, for SNBN the value of $T_m-T_{VF}$ difference (19 K, at 1 kHz) is much smaller and it is comparable to that of $Pb(Mg_{1/3}Nb_{2/3})O_3$ (23 K).[18]

Relaxor-like behavior occurs mainly because of the local compositional fluctuations and/or structural disorder in the arrangement of cations in one or more crystallographic sites of the structure.[12,20] However, many Aurivillius phases have atomic positions with a mixed occupancy on the A and/or B site of the perovskite blocks but without being associated with a relaxor behavior.[14,15] For SNBN, $Na^+$ and $Sr^{2+}$ occupy the A-site of the perovskite blocks in 50:50 ratio. Disorder at the A-site of perovskite blocks due to the random distribution of $Sr^{2+}$ and $Na^+$ seems to be responsible for observed relaxor behavior in SNBN. A-site disorder and associated relaxor properties have been observed in perovskite $Na_{0.5}Bi_{0.5}TiO_3$, as well as in two-layered Aurivillius phases $K_{0.5}La_{0.5}Bi_2(Nb/Ta)_2O_9$.[7,8,21]

Room temperature (295 K) P-E loop recorded at a frequency of 50 Hz for SNBN ceramics is shown in Fig. 4(a). A remnant polarization ($2P_r$) of 4.1 µC/cm$^2$, maximum polarization ($2P_{max}$) of 15.1 µC/cm$^2$ and a coercive field ($2E_C$) of 28.3 kV/cm were obtained under a maximum applied electric field of 88 kV/cm. Unlike normal ferroelectrics for which polarization becomes zero at the Curie temperature $T_C$, relaxor ferroelectrics are known to possess non-zero polarization at temperatures well above $T_m$ which is attributed to the presence of micro/nanoscopic polar regions.[12] It is widely believed that these polar nano/micro regions are responsible for the relaxor behavior. Several models have been proposed for the mechanism of formation of microscopic polar regions.[12,18,21-31] Formation of these polar regions occurs at $T_B$, so called Burns temperature (analogue of the freezing of the spin fluctuations in a magnetic system), which could be hundreds of degrees higher than $T_m$.[31] At temperature close to $T_B$, these polar regions are in ergodic state. Upon cooling, dynamics of polar regions slow down and finally freeze at $T_{VF} < T_m$.[22] To corroborate the presence of spontaneous polarization above the phase transition, P-E loops recorded at three different temperatures (above $T_m$) are shown in Figs. 4 (b)-(d). At these temperatures P-E loop becomes pinched (or constricted). Pinched P-E loops have also been observed in compounds with antiferroelectric component and also in doped-ferroelectrics after ageing.[32-34] This phenomenon in accepter-doped ferroelectrics is considered to relate to defect dipoles providing a restoring force to polarization reversal, thus 'pinning' the motion of ferroelectric domains.[33] However, ageing related phenomena do not seem to be responsible for the constricted loops observed in our sample. While it is true that some amount of oxygen vacancies are generally present in Aurivillius family of oxides due to volatilization of bismuth during high temperature sintering but for the present sample pinching of hysteresis loop is more pronounced above $T_m$ whereas no pinching is apparent in the hysteresis loop recorded at 295 K (Fig. 4). Reversibility in temperature dependent shape of P-E loops indicates that such a behavior in SNBN might arise due to the coexistence of polar and non-polar regions at these temperatures.

**IV Conclusions**

We have investigated the structural and relaxor behavior of $SrNaBi_2Nb_3O_{12}$ ceramics. The Rietveld refinement of X-ray powder data suggests that SNBN crystallizes in the orthorhombic space group *B2cb*. SNBN ceramics exhibited near-room temperature frequency-dependent dielectric maximum which follows the Vogel-Fulcher relation implying a relaxor nature. The relaxor behavior seems to have its origin in inhomogeneous distribution of $Na^+$ and $Sr^{2+}$ at A-site. Pinched P-E loops were observed well above $T_m$.

**Acknowledgements**

One of the authors (S.K.) is grateful to the Council of Scientific and Industrial Research (CSIR), New Delhi, for the award of Senior Research Fellowship. S.K. also thanks the European Commission for providing fellowship through EMECW-WILLpower programme to carry out some of the present investigations at UPC, Barcelona, Spain.

**Figure captions:**

Figure 1. (a) Observed (circles), calculated (solid line), and difference profile (bottom line) obtained from the Rietveld refinement of X-ray powder diffraction profile for $SrNaBi_2Nb_3O_{12}$ using $B2cb$ space group. The vertical bars are the Bragg positions of the reflections in the space group $B2cb$. (b) High resolution transmission electron micrograph recorded along the [100] zone axis. (c) Scanning electron micrograph of fractured surface of the ceramic sintered at 1323 K for 4 h.

Figure 2. Temperature dependence of real ($\varepsilon_r'$) and imaginary ($\varepsilon_r''$) parts of dielectric constant at different frequencies.

Figure 3. Temperature of dielectric maximum $T_m$ as a function of ln ($f$) (*spheres* represent the experimental data and the *solid line* is the fit to the Vogel–Fulcher relation).

Figure 4. (a) Polarization vs Electric field hysteresis loop recorded at 295 K, (b) 330 K, (c) 350 K, and (d) 370 K for SNBN ceramics.

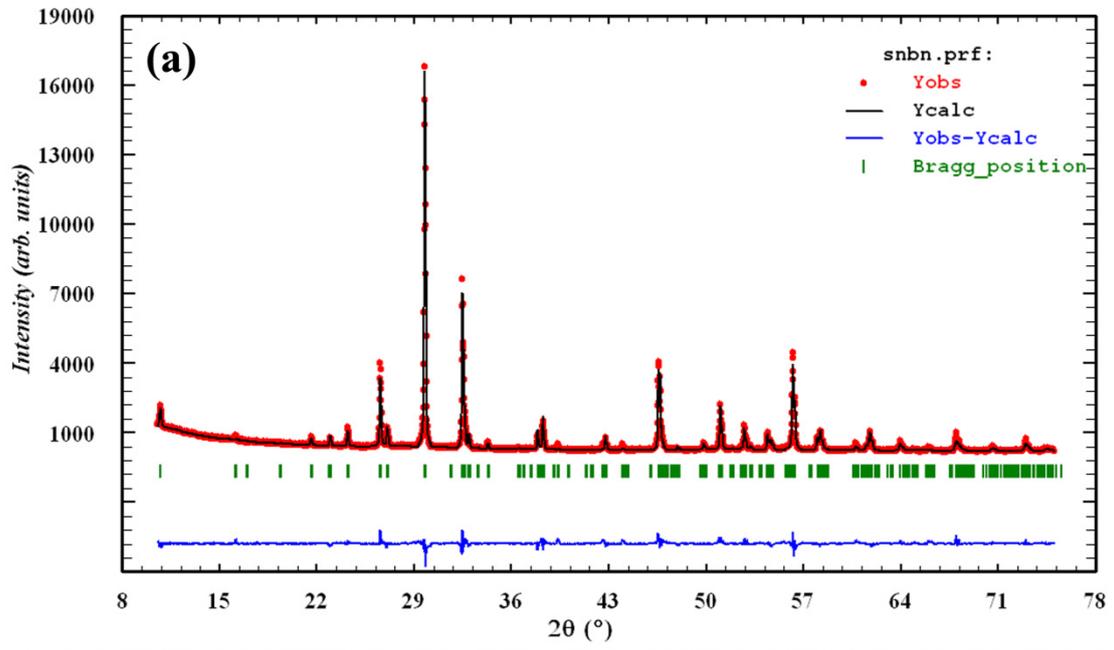

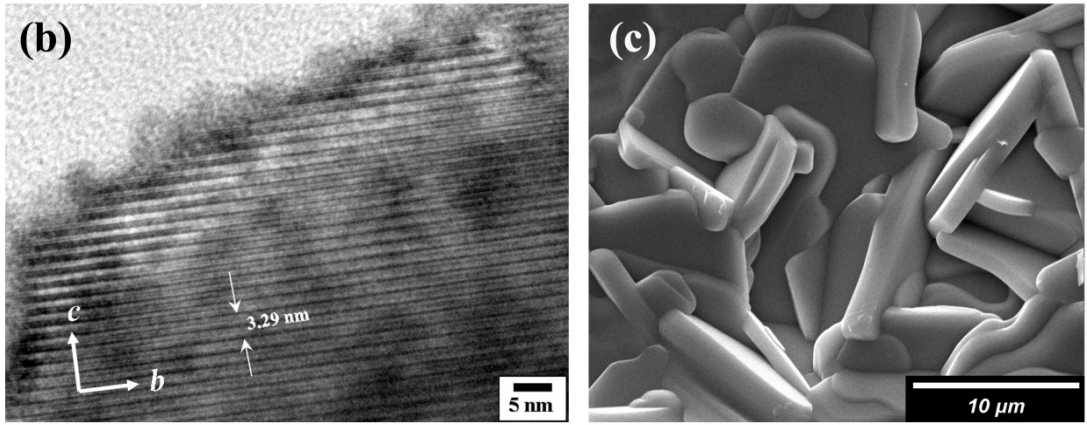

*Figure 1*

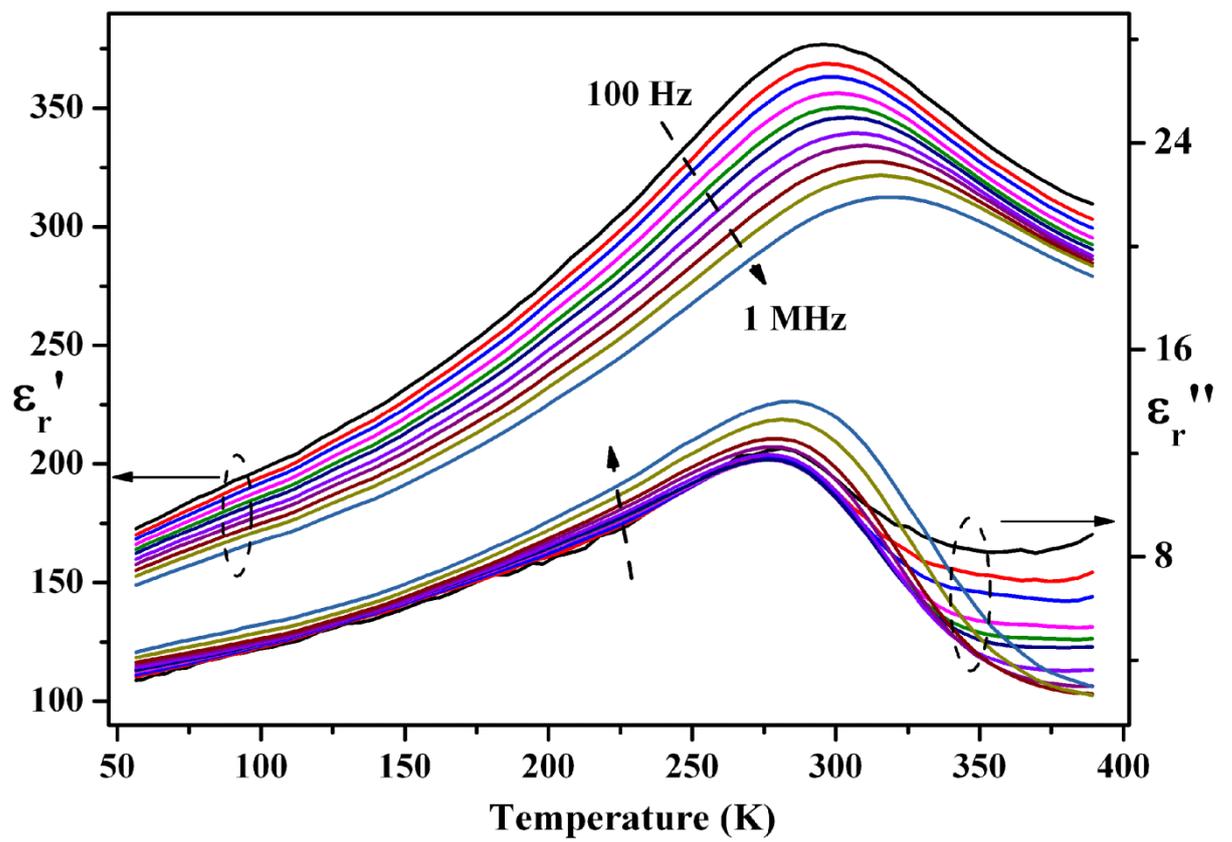

*Figure 2*

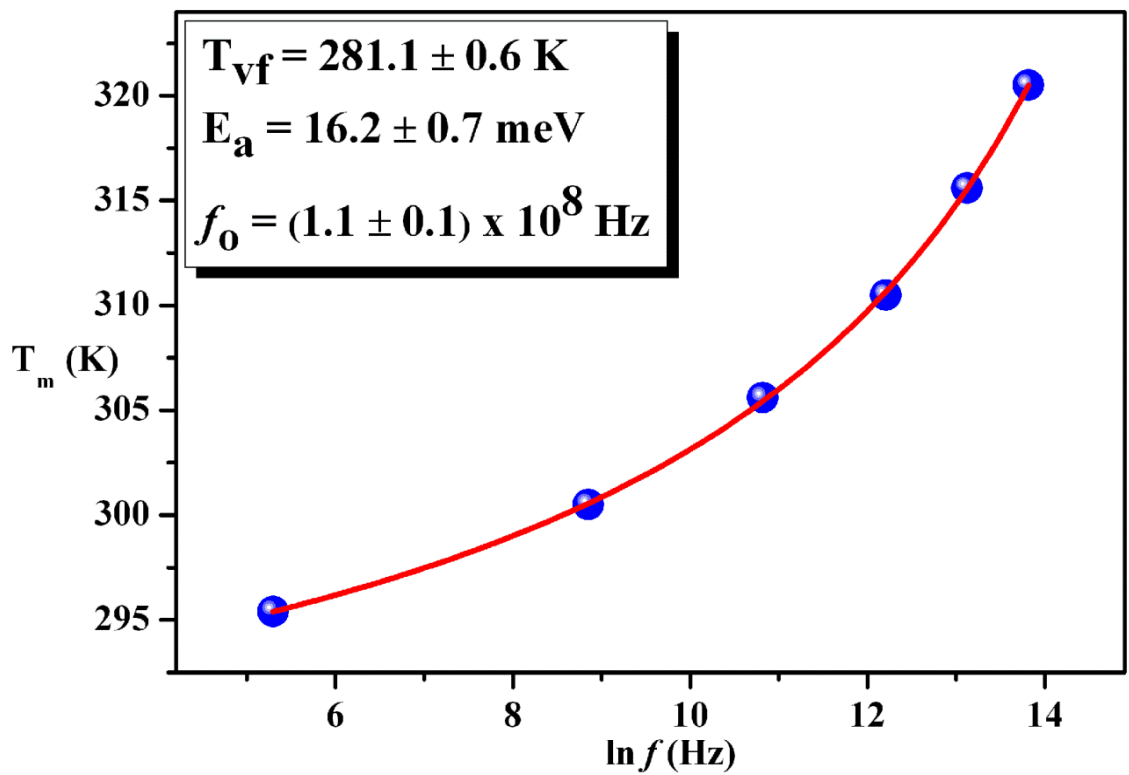

$T_{vf}$ = 281.1 ± 0.6 K

$E_a$ = 16.2 ± 0.7 meV

$f_0$ = (1.1 ± 0.1) x $10^8$ Hz

$T_m$ (K)

ln $f$ (Hz)

*Figure 3*

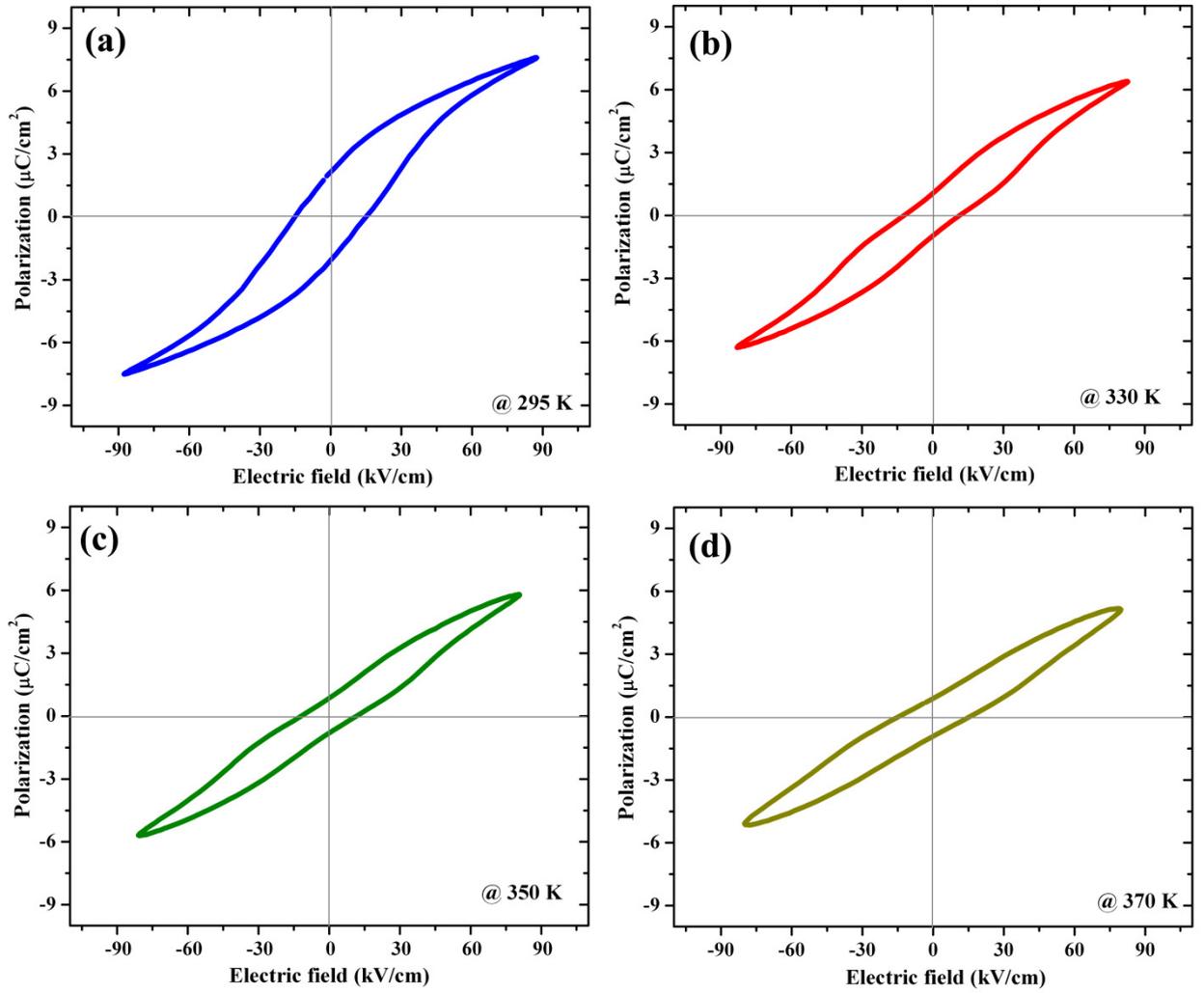

*Figure 4*